\documentclass[conference]{IEEEtran}
\usepackage[T1]{fontenc}
\usepackage[utf8]{inputenc}
\usepackage{csquotes}
\usepackage{mathtools,amssymb,amsthm}
\usepackage{nicefrac}
\makeatletter
\let\MYcaption\@makecaption
\makeatother
\usepackage{subcaption}
\makeatletter
\let\@makecaption\MYcaption
\makeatother
\usepackage{booktabs}
\usepackage[noadjust]{cite}
\usepackage[binary-units=true]{siunitx}
\usepackage{tikz}
\usepackage[textsize=tiny]{todonotes}
\usepackage[bookmarks=false]{hyperref}

\usetikzlibrary{arrows.meta,backgrounds,calc,matrix,positioning,fit}

\newtheorem{example}{Example}
\newcommand{\ket}[1]{\ensuremath{\left|#1\right\rangle}}
\newcommand{\iu}{\mathrm{i}\mkern1mu}

\hypersetup{
	pdftitle={Just Like the Real Thing: Fast Weak Simulation of Quantum Computation},
	pdfsubject={Design Automation Conference 2020},
	pdfauthor={Stefan Hillmich, Igor L. Markov, and Robert Wille}
}

\begin{document}
\title{\Huge Just Like the Real Thing:\\Fast Weak Simulation of Quantum Computation}
\author{Stefan Hillmich$^1$, Igor L. Markov$^2$, and Robert Wille$^1$ \\ 
	{\normalsize $^1$Institute for Integrated Circuits, Johannes Kepler University Linz, Austria}\\
	{\normalsize $^2$Department of EECS, University of Michigan, USA}\\
	{\normalsize stefan.hillmich@jku.at, imarkov@eecs.umich.edu, robert.wille@jku.at}\\
	{\small\url{https://iic.jku.at/eda/research/quantum/}}}
\date{}
\maketitle
\begin{abstract}
	Quantum computers promise significant speedups in solving problems intractable for conventional computers but, despite recent progress, remain limited in scaling and availability. Therefore, quantum software and hardware development heavily rely on simulation that runs on conventional computers. Most such approaches perform \emph{strong simulation} in that they explicitly compute amplitudes of quantum states. However, such information is not directly observable from a physical quantum computer because quantum measurements produce random samples from probability distributions defined by those amplitudes. In this work, we focus on \emph{weak simulation} that aims to produce outputs which are statistically indistinguishable from those of error-free quantum computers.
    We develop algorithms for weak simulation based on quantum state representation in terms of decision diagrams. We compare them to using state-vector arrays and binary search on prefix sums to perform sampling.
    Empirical validation shows, for the first time, that this enables mimicking of physical quantum computers of significant scale.
\end{abstract}

\vspace*{-1em}
	
\section{Introduction}

Quantum computing \cite{NC:2000} promises to fundamentally change the field of computing and its applications. For example, Shor's algorithm~\cite{Sho:94} performs integer factorization in low polynomial time and poses a severe threat to modern cryptography which relies on the hardness of integer factorization.
Applications proposed more recently include
search for better catalysts in quantum chemistry~\cite{cao2018quantum}, as well as
machine learning, cryptography, quantum simulation, and solving systems of linear equations~\cite{montanaro2016quantum,preskill2018quantum,coles2018quantum}.
Their potential has been recognized by Google, IBM, Microsoft as well as start-ups such as
Rigetti and IonQ which heavily invest in this technology.

Despite initial optimism, the construction of quantum computers and the implementation of quantum algorithms turned out exceptionally challenging. Quantum computers available today
are expensive, error-prone, limited in their scalability, and inaccessible to most researchers.
Therefore, simulation methods which faithfully mimic the behavior of a quantum computer on conventional hardware are essential to the design, optimization, verification, and performance evaluation
of quantum algorithms and their applications. However, \mbox{state-of-the-art} techniques for quantum circuit simulation~\mbox{\cite{jones2019quest,DBLP:journals/corr/WeckerS14,qxSimulator2017,DBLP:journals/corr/SmelyanskiySA16,VMP:2009,zulehner2017advanced,DBLP:journals/qip/WangHH17}} remain somewhat disconnected from this goal because they
primarily focus on so-called \emph{strong simulation}, i.e.,~explicitly compute some or all of the amplitudes of the final quantum state produced by a given circuit. Notably, such amplitudes cannot be
directly observed from a quantum computer. Instead, every run of a quantum computer produces \mbox{nondeterministic} outputs of quantum measurements which can be interpreted as indices (represented
in binary) of amplitudes and sampled according to probabilities computed as squared norms of these amplitudes.
In contrast to \emph{strong simulation}, the task of mimicking such nondeterministic output, possibly with some error, is called \emph{weak simulation}.
The two tasks are not equivalent, but weak simulation has not yet received extensive coverage~\cite{nest2008classical,Bravyi_2019}.

In this work, we develop fast methods for weak simulation. Since there often is no need to return
exponential numbers of amplitudes, we (1) represent quantum states using a compressed data structure called \emph{edge-weighted decision diagram}, and (2) develop novel algorithms to simulate measurements in terms of this data structure in \emph{linear time} with respect to its size.
To validate our approach and provide a performance baseline, we also show how to simulate measurement on an explicit array of all amplitudes by using prefix sums and binary search. Empirical validation confirms that the reduced memory needs of decision diagrams help making sampling from quantum states more practical. In comparison, state representations that use full arrays require exorbitant amounts of memory
and often cannot perform weak simulation in practice. Our empirical results demonstrate, for the first time,
that physical quantum computers running several well-known quantum algorithms on
a significant scale can be mimicked faithfully and efficiently by simulators running on modest conventional computers.

The rest of the paper is structured as follows:
Section~\ref{sec:background} reviews the background on quantum states and operations,
so as to make the paper accessible to readers without a keen understanding of quantum computing.
Section~\ref{sec:weak-simulation} outlines the main idea of using weak simulation to mimic a quantum computer and also describes algorithms that rely on \mbox{exponentially-sized} arrays. All concepts and algorithms are illustrated by detailed examples.
Section~\ref{sec:advanced-ws} introduces our approach to weak simulation
without exponentially-sized arrays. This section reviews how decision diagrams represent quantum states, then describes our algorithm for weak simulation and an
enhancement for decision diagrams in this context.
Empirical validation of our implementation is described in Section~\ref{sec:evaluation}.
Section~\ref{sec:conclusion} concludes the paper.

\section{Background: Quantum Computing}
\label{sec:background}

In the realm of quantum computing, classical bits are generalized to \emph{quantum bits} or \emph{qubits}.
While the former can be either in state \(0\) or in state \(1\), qubits may assume one of two basis states (denoted~\(\ket{0}\) and \(\ket{1}\)) and also any linear combination of them. This is described by \( \ket{\psi} = \alpha_0\cdot\ket{0} + \alpha_1\cdot\ket{1} \) with \emph{amplitudes} \(\alpha_0, \alpha_1 \in \mathbb{C} \) which have to satisfy the normalizing constraint \( |\alpha_0|^2 + |\alpha_1|^2 = 1 \).
Qubits with \(\alpha_0\neq 0\) and \(\alpha_1\neq 0\) are said to be in \emph{superposition}.\footnote{Another important phenomenon is \emph{entanglement}, where the measurement of a single qubit may influence the measured result of another qubit.}

For multi-qubit quantum systems, the description is extended accordingly to represent the exponential number of basis states the systems can assume. For example, a system with two qubits has four basis states, i.e.,~\mbox{\( \ket{\psi} = \alpha_{00}\cdot\ket{00} + \alpha_{01}\cdot\ket{01} + \alpha_{10}\cdot\ket{10} + \alpha_{11}\cdot\ket{11} \)}. Since the amplitudes dictate the probabilities, the normalizing constraint is extended as well: \mbox{\( |\alpha_{00}|^2 + |\alpha_{01}|^2 + |\alpha_{10}|^2 + |\alpha_{11}|^2 = 1 \)}.
Commonly, the description of a quantum state is shortened to a vector containing only the amplitudes, e.g.,~\( \ket{\psi} = \left[\alpha_{00}, \alpha_{01}, \alpha_{10}, \alpha_{11}\right]^\mathrm{T} \).

\begin{example}
	\label{ex:qubits}
	Consider an arbitrary quantum system composed of two qubits, which is in the state
	\mbox{\(\ket{\psi} = \nicefrac{1}{\sqrt{2}}\cdot\ket{00} + 0\cdot\ket{01} + 0\cdot\ket{10} +\nicefrac{1}{\sqrt{2}}\cdot\ket{11}\)}.
	This represents a valid state, since \({\left|\nicefrac{1}{\sqrt{2}}\right|^2 + 0^2 + 0^2 + \left|\nicefrac{1}{\sqrt{2}}\right|^2 = 1}\).
	The corresponding state vector is
	\(
	\ket{\psi} = \left[ \nicefrac{1}{\sqrt{2}}, 0, 0, \nicefrac{1}{\sqrt{2}} \right]^\mathrm{T}.
	\)
	
\end{example}

Before quantum measurement is applied, the state of a quantum system can be manipulated using unitary quantum operations.
Such operations are defined through unitary matrices, i.e.,~square matrices whose inverse is their conjugate transposed~\cite{NC:2000}.
Examples of important single-qubit quantum operations are
\[
	X= \begin{bmatrix} 0 & 1 \\ 1 & 0 \end{bmatrix},
	H = \frac{1}{\sqrt{2}}\begin{bmatrix*}[r] 1 & 1 \\ 1 & -1 \end{bmatrix*} \text{, and }
	Z = \begin{bmatrix*}[r] 1 & 0 \\ 0 & -1 \end{bmatrix*},
\]
where $X$ negates the state of the qubit, $H$ sets the qubit into superposition, and $Z$ shifts the phase of the qubit. To couple multiple qubits, one can use, e.g.,~the $\mathit{CNOT}$ (controlled-\(\mathit{NOT}\)) operation, which negates a \emph{target qubit}, iff the chosen \emph{control qubit} is in the state \ket1. This is defined through the matrix
\[
	\mathit{CNOT} = \begin{bmatrix} 1 & 0 & 0 & 0 \\ 0 & 1 & 0 & 0 \\ 0 & 0 & 0 & 1 \\ 0 & 0 & 1 & 0 \end{bmatrix}.
\]

The action of a quantum operation represented by a matrix on a quantum state represented by a vector can be described through matrix-vector multiplication as illustrated next:
\begin{example}\label{ex:qua_op}
	Consider a quantum system composed of two qubits which is currently in state \(\ket{\psi} = \ket{00}\). Performing an \(H\) operation on the first qubit and a \(\mathit{CNOT}\) operation (with control on the first qubit and target on the second) yields a new state \(\ket{\psi'}\) determined by multiplying the vector~$\ket{\psi}$ with the matrices of these two operations, i.e.,
	\[
	\underbrace{\frac{1}{\sqrt{2}}\begin{bmatrix*}[r]
		1 & 0 & 1 & 0 \\
		0 & 1 & 0 & 1 \\
		1 & 0 & -1 & 0 \\
		0 & 1 & 0 & -1 \\
		\end{bmatrix*}}_{\text{\(H\) on \(\text{1}^\text{st}\) qubit}}
	\times
	\underbrace{\begin{bmatrix} 1 & 0 & 0 & 0 \\ 0 & 1 & 0 & 0 \\ 0 & 0 & 0 & 1 \\ 0 & 0 & 1 & 0 \end{bmatrix}}_{\mathit{CNOT}}
	\times
	\underbrace{\begin{bmatrix} 1 \\ 0 \\ 0 \\ 0 \end{bmatrix}}_{\ket{\psi}}
	=
	\underbrace{\frac{1}{\sqrt{2}}\begin{bmatrix} 1 \\ 0 \\ 0 \\ 1 \end{bmatrix}}_{\ket{\psi'}}.
	\]
\end{example}

Unfortunately, the amplitudes \(\alpha_{00}\), \(\alpha_{01}\), \(\alpha_{10}\), and \(\alpha_{11}\) of the resulting output state cannot be observed directly on a quantum computer. Instead, these amplitudes dictate the probability of certain outcomes of a measurement with respect to the corresponding basis states. More precisely, measuring a single qubit in state
\mbox{\( \ket{\psi} = \alpha_{00}\cdot\ket{00} + \alpha_{01}\cdot\ket{01} + \alpha_{10}\cdot\ket{10} + \alpha_{11}\cdot\ket{11} \)}
yields the output \(\ket{00}\) with probability~\( |\alpha_{00}|^2 \), the output~\(\ket{01}\) with probability~\( |\alpha_{01}|^2 \), etc.
After the measurement, the qubits will lose any superposition, i.e., they collapse into a basis state.

\begin{example}
Consider again the state \(
\ket{\psi'} = \left[ \nicefrac{1}{\sqrt{2}}, 0, 0, \nicefrac{1}{\sqrt{2}} \right]^\mathrm{T}
\)
produced in Example~\ref{ex:qua_op}.
Measuring its qubits does \emph{not} provide the amplitudes of this state, but only yields one of the possible basis states (here: \(\ket{00}\) or \(\ket{11}\); both with probability of \mbox{\(|\alpha_{00}|^2=|\alpha_{11}|^2=\left| \nicefrac{1}{\sqrt{2}}\right|^2 = \nicefrac{1}{2}\)}).
After the measurement, both qubits will lose their superposition, i.e., the state collapses to the resulting basis state.
\end{example}

A common way to represent quantum computations is with quantum circuit diagrams~\cite{NC:2000}. Here, the qubits are represented by horizontal lines, while quantum operations are placed on the qubits and are applied from left to right.

\begin{figure}
	\centering
	\begin{tikzpicture}[gate/.style={draw,fill=white,minimum size=1.5em},
	control/.style={draw,fill,shape=circle,minimum size=5pt,inner sep=0pt},
	cross/.style={fill=white,path picture={\draw[thick,black](path picture bounding box.north) -- (path picture bounding box.south) (path picture bounding box.west) -- (path picture bounding box.east);}},
	target/.style={draw,circle,cross,minimum width=0.3cm},
	classic/.style={double, double distance=1.25pt},
	meter/.append style={
		draw,
		rectangle,
		minimum size=1.5em,
		fill=white,
		path picture={
			\draw[black] ([shift={(1pt,5pt)}]path picture bounding box.south west) to[bend left=50] ([shift={(-1pt,5pt)}]path picture bounding box.south east);
			\draw[black,-latex] ([shift={(0,.1)}]path picture bounding box.south) -- ([shift={(.3,-.1)}]path picture bounding box.north);}},
	x=0.75cm, y=0.75cm]
	\node[anchor=east] (q0) at (0.25,2) {$\ket{q_0}$};
	\node[anchor=east] (q1) at (0.25,1) {$\ket{q_1}$};
	\node[anchor=east] (q2) at (0.25,0) {$\ket{q_2}$};
	
	\node[gate] at (1,0) {\(H\)};
	\draw (2,1) node[control] {} -- (2,0) node[target] {};
	\node[gate] at (3,1) {\(X\)};
	\draw (4,1) node[control] {} -- (4,2) node[target] {};
	
	\node[meter] (meter0) at (5, |-q0) {};
	\node[meter] (meter1) at (5, |-q1) {};
	\node[meter] (meter2) at (5, |-q2) {};
	
	\begin{scope}[on background layer]
	\draw (q2) -- (meter2) edge[classic] ++(0.75,0);
	\draw (q1) -- (meter1) edge[classic] ++(0.75,0);
	\draw (q0) -- (meter0) edge[classic] ++(0.75,0);
	\end{scope}
	\end{tikzpicture}
	\caption{A quantum circuit diagram}
	\label{fig:qantum-circuit}
	\vspace*{-1em}
\end{figure}
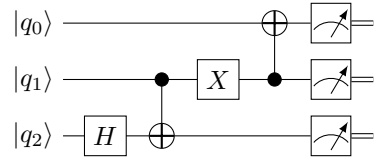

\begin{example}
	Fig.~\ref{fig:qantum-circuit} shows a quantum circuit diagram with four operations: \(H\), \(\mathit{CNOT}\) (\(\bullet\) as control, \(\oplus\) as target), \(X\), and another \(\mathit{CNOT}\), followed by a measurement on each qubit.
	The double lines after the measurements indicate that the qubit is in a basis state here (as a result of the measurement).
\end{example}

\section{Weak Simulation:\\Mimicking Physical Quantum Computers}
\label{sec:weak-simulation}

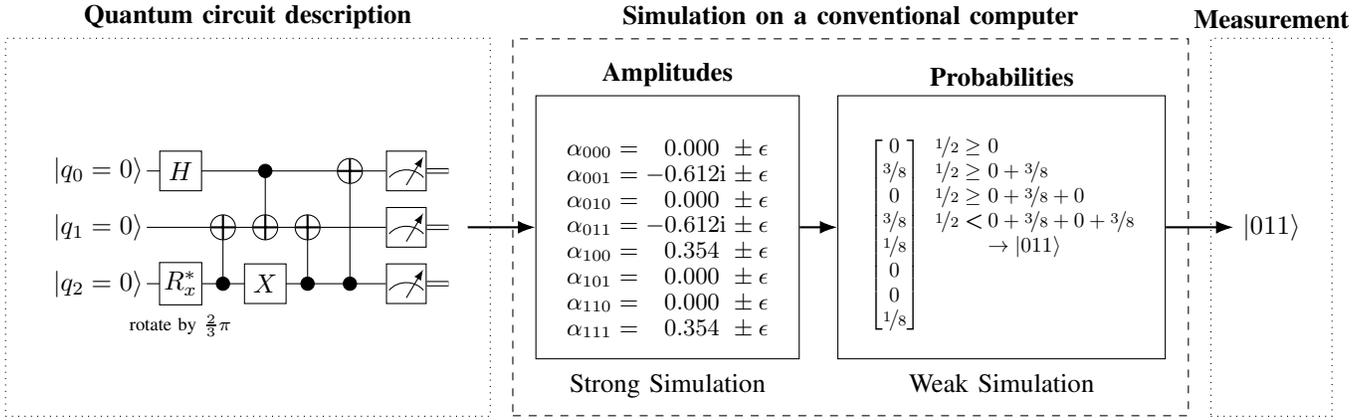
\begin{figure*}[tbp]
	\centering
	\newsavebox\plotbox
	\begin{lrbox}{\plotbox}
		\begin{minipage}{0.3\linewidth}%
		\begin{align*}
			\begin{bmatrix*}
				0 \\
				\nicefrac{3}{8} \\
				0 \\
				\nicefrac{3}{8} \\
				\nicefrac{1}{8} \\
				0 \\
				0 \\
				\nicefrac{1}{8}
			\end{bmatrix*}
			\enskip
			\setlength\arraycolsep{1pt}\begin{matrix*}[l]
				\nicefrac{1}{2} \ge &0  \\
				\nicefrac{1}{2} \ge &0 + \nicefrac{3}{8} \\
				\nicefrac{1}{2} \ge &0 + \nicefrac{3}{8} + 0 \\
				\nicefrac{1}{2} \boldsymbol{<} &0 + \nicefrac{3}{8} + 0 + \nicefrac{3}{8} \\
									& \to \ket{011}\\
				\\
				\\
				\\
			\end{matrix*}
		\end{align*}
		\end{minipage}
	\end{lrbox}
	\newsavebox\circuitbox
	\begin{lrbox}{\circuitbox}
			\begin{tikzpicture}[gate/.style={draw,fill=white,minimum size=1.5em},
				control/.style={draw,fill,shape=circle,minimum size=5pt,inner sep=0pt},
				cross/.style={fill=white,path picture={\draw[thick,black](path picture bounding box.north) -- (path picture bounding box.south) (path picture bounding box.west) -- (path picture bounding box.east);}},
				target/.style={draw,circle,cross,minimum width=0.3cm},
				classic/.style={double, double distance=1.25pt},
				meter/.append style={
					draw,
					rectangle,
					minimum size=1.5em,
					fill=white,
					path picture={
						\draw[black] ([shift={(1pt,5pt)}]path picture bounding box.south west) to[bend left=50] ([shift={(-1pt,5pt)}]path picture bounding box.south east);
						\draw[black,-latex] ([shift={(0,.1)}]path picture bounding box.south) -- ([shift={(.3,-.1)}]path picture bounding box.north);}},
				x=0.75cm, y=0.75cm]
				\node[anchor=east] (q0) at (0,2) {\(\ket{q_0 = 0}\)};
				\node[anchor=east] (q1) at (0,1) {\(\ket{q_1 = 0}\)};
				\node[anchor=east] (q2) at (0,0) {\(\ket{q_2 = 0}\)};
				
				\node[gate,label={[overlay]below:\scriptsize rotate by \( \frac{2}{3}\pi \)}] at (0.5, |-q2) {\!\(R_x^*\)\!}; 
				\node[gate] at (0.5, |-q0) {\(H\)};
				
				\draw (1.25,|-q1) node[target] {} -- (1.25,|-q2) node [control] {};
				\node[gate] at (2.00, |-q2) {\(X\)};
				
				\draw (2,|-q0) node[control] {} -- (2,|-q1) node [target] {};
				\draw (2.75,|-q2) node[control] {} -- (2.75,|-q1) node [target] {};
				\draw (3.5,|-q2) node[control] {} -- (3.5,|-q0) node [target] {};
				
				\node[meter] (meter0) at (4.5, |-q0) {};
				\node[meter] (meter1) at (4.5, |-q1) {};
				\node[meter] (meter2) at (4.5, |-q2) {};
				
				\begin{scope}[on background layer]
					\draw (-0.1,0) -- (meter2) edge[classic] ++(0.75,0);
					\draw (-0.1,1) -- (meter1) edge[classic] ++(0.75,0);
					\draw (-0.1,2) -- (meter0) edge[classic] ++(0.75,0);
				\end{scope}
			\end{tikzpicture}
	\end{lrbox}
	\begin{tikzpicture}
		\node[minimum width=3.75cm, minimum height=3.5cm, draw, inner sep=0pt] (weak) {	
			\scalebox{.8}{\usebox\plotbox}
		};
		\node[right = 0.9of weak, align=center] (output) {\ket{011}};
		
		\node[left =0.5 of weak, draw, text width=3.25cm, minimum width=3.5cm, minimum height=3.5cm] (strong) {
			\centering
			\small
			\setlength{\jot}{0pt}
			\begin{align*}
				\alpha_{000} &= \phantom{-}0.000\phantom{\iu} \pm \epsilon \\
				\alpha_{001} &= -0.612\iu \pm \epsilon \\
				\alpha_{010} &= \phantom{-}0.000\phantom{\iu} \pm \epsilon \\
				\alpha_{011} &= -0.612\iu \pm \epsilon \\
				\alpha_{100} &= \phantom{-}0.354\phantom{\iu} \pm \epsilon \\
				\alpha_{101} &= \phantom{-}0.000\phantom{\iu} \pm \epsilon \\
				\alpha_{110} &= \phantom{-}0.000\phantom{\iu} \pm \epsilon \\
				\alpha_{111} &= \phantom{-}0.354\phantom{\iu} \pm \epsilon
			\end{align*}\par
		};
		
		\node[left=0.9 of strong, inner sep=6pt] (input) {\usebox\circuitbox};
		
		\node[above = 0cm of weak, font=\bfseries] (weaklabel) {Probabilities};
		\node[below = 0cm of weak, yshift=-2pt] (weaklabel2) {Weak Simulation};
		
		\node[above = 0cm of strong, font=\bfseries] (stronglabel) {Amplitudes};
		\node[below = 0cm of strong, yshift=-2pt] (stronglabel2) {Strong Simulation};
		
		\draw[-Latex, thick] (weak.east) -- (output.west);
		\draw[-Latex, thick] (strong.east) -- (weak.west);
		\draw[-Latex, thick] (input.east) -- (strong.west);
		
		\node[fit=(strong)(weak), inner xsep=0.3cm, inner ysep=0.75cm, draw, dashed, label={[above,font=\bfseries]:Simulation on a conventional computer}] (simulation) {};
		
		\node[fit=(input.south east |- simulation.south east)(input.north west |- simulation.north west), inner xsep=0.3cm, inner ysep=0, draw, dotted, label={[above,font=\bfseries]:Quantum circuit description}] (inputbox) {};
		
		\node[fit=(output.south east |- simulation.south east)(output.north west |- simulation.north west), inner xsep=0.3cm, inner ysep=0, draw, dotted, label={[above,font=\bfseries]:Measurement}] (samplebox) {};
	\end{tikzpicture}
	\vspace{-0.75em}
	\caption{Mimicking a physical quantum computer by generating individual output samples after strong simulation}
	\label{fig:quantum_computer_simulation}
	\vspace*{-1.5em}
\end{figure*}

In this work, we explore the simulation of quantum computers on conventional hardware
that produces the same kind of output as the physical quantum computers. Moreover, we hope to produce outputs that are statistically indistinguishable from those of (error-free) physical quantum computers.
Given the well-defined model of quantum computation covered in Section \ref{sec:background},
this task seems
straightforward at a first glance: We are given an input basis state and a sequence of quantum operations,
and need to (nondeterministically) produce bitstrings that represent measurement outcomes.
However, performing this kind of simulation without unnecessary overhead has been elusive,
and the literature focuses on explicitly describing the output distribution rather than producing
individual output samples efficiently~\cite{nest2008classical,Bravyi_2019}.
In this section, we describe and illustrate relevant challenges using the following example:

\begin{example}
	Consider Fig.~\ref{fig:quantum_computer_simulation}~(left) which illustrates
	the running example used in the remainder of this work.
	The dotted box specifies the input basis state \ket{000}
	along with a sequence of quantum operations (given as a quantum circuit diagram)
	to be simulated.
\end{example}

Based on the $n$-qubit input state and circuit operations,
the $2^n$ amplitudes \( \alpha_i \) of the corresponding output state
can be calculated by matrix-vector multiplication (see Section \ref{sec:background}).

\begin{example}
	Using the vector representation of the input basis state \ket{000} and the matrix representations of circuit operations, we can determine the output state vector through a series of matrix-vector multiplications by calculating the amplitudes \( \alpha_{000}, \alpha_{001}, \alpha_{010}, \ldots, \alpha_{111} \), which cannot be produced by one run of a quantum computer. As indicated in Fig.~\ref{fig:quantum_computer_simulation}~(middle), computing those amplitudes with a small error is also
acceptable.
\end{example}

Finally, to simulate a quantum measurement in the computational basis,
we sample an output-basis state from the probability distribution $(p_i)$
where $p_i = \alpha_i^*\alpha_i^{\phantom{*}} = |\alpha|^2$.

\begin{example}
	Fig.~\ref{fig:quantum_computer_simulation}~(right) illustrates a probability distribution
	of measurement outcomes. Sampling from that distribution may yield \ket{001}.
\end{example}

To faithfully mimic a physical quantum computer, one can first capture the output distribution
$(p_i)$ via strong simulation and, then, sample from it, as outlined in Fig.~\ref{fig:quantum_computer_simulation}. If the output probability distribution
is described explicitly by a vector of probabilities $p_i$, sampling can be performed
using a standard \emph{biased random selection} routine. The idea is to generate
a random number \(\hat{p} \in [0,1)\) and, then, determine the largest
index $i$ such that $\Sigma_{k=0}^i p_k \leq \hat{p}$.

\begin{example}
	Continuing our running example in Fig.~\ref{fig:quantum_computer_simulation}~(right), we find the probability \(p_i\) of each basis state. Now, assuming that \(\hat{p}=\nicefrac{1}{2}\) was generated randomly, the fourth prefix sum exceeds \(\hat{p}\) since \(0 + \nicefrac{3}{8} + 0 + \nicefrac{3}{8} > \nicefrac{1}{2}\). Therefore, $i=3$ and the resulting sample is \(\ket{011}\).
\end{example}

\begin{figure}
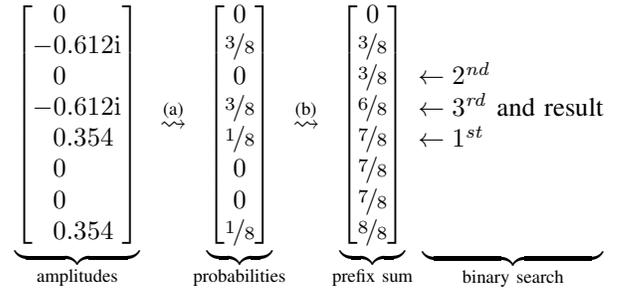

	\[
	\underbrace{\begin{bmatrix*}[l]
		\phantom{-}0 \\
		-0.612\iu \\
		\phantom{-}0 \\
		-0.612\iu \\
		\phantom{-}0.354 \\
		\phantom{-}0 \\
		\phantom{-}0 \\
		\phantom{-}0.354
		\end{bmatrix*}}_{\text{\makebox[0pt]{amplitudes}}}
	\enskip\overset{\textup{(a)}}{\rightsquigarrow}\enskip
	\underbrace{\begin{bmatrix*}
		0 \\
		\nicefrac{3}{8} \\
		0 \\
		\nicefrac{3}{8} \\
		\nicefrac{1}{8} \\
		0 \\
		0 \\
		\nicefrac{1}{8}
		\end{bmatrix*}}_{\text{\makebox[0pt]{probabilities}}}
	\enskip\overset{\textup{(b)}}{\rightsquigarrow}\enskip
	\underbrace{\begin{bmatrix*}
		0 \\
		\nicefrac{3}{8} \\
		\nicefrac{3}{8} \\
		\nicefrac{6}{8} \\
		\nicefrac{7}{8} \\
		\nicefrac{7}{8} \\
		\nicefrac{7}{8} \\
		\nicefrac{8}{8}
		\end{bmatrix*}}_{\text{\makebox[0pt]{prefix sum}}}
	\enskip\underbrace{\begin{matrix*}[l]
		\\
		\\
		\!\leftarrow 2^\mathit{nd}\\
		\!\leftarrow 3^\mathit{rd} \text{ and result}\\
		\!\leftarrow 1^\mathit{st}\\
		\\
		\\
		\\
		\end{matrix*}}_{\text{\makebox[0pt]{binary search}}}
	\]
	\vspace*{-1.25em}
	\caption{Biased random selection via binary search on a prefix array. The precomputation in (a) is performed once to facilitate efficient repeated sampling in (b).}
	\label{fig:advanced_weak_sim}
\vspace*{-1.5em}
\end{figure}

The index $i$ can be found by a direct (linear) traversal, which takes $2^{n-1}$ steps on average. To accelerate \emph{repeated sampling}, one can first compute the prefix values $r_i=\Sigma_{k=0}^i p_k$ and store them in an array, noting
that $r_i \leq r_{i+1}$ for all $i$. Then, for each newly-generated $\hat{p}$,
find $i$ using binary search in \mbox{$\mathcal{O}(\log 2^n)=\mathcal{O}(n)$} time.
Fig.~\ref{fig:advanced_weak_sim} illustrates the computation of the prefix array and subsequent sampling by binary search in this array.

The use of precomputation and repeated binary search makes it possible to draw a large number of samples much more efficiently than with linear traversals. However, the use of binary search requires random access to the prefix array, which must be loaded in memory. In contrast, linear traversals can be performed on large vectors stored in out-of-memory files, with only small blocks loaded to memory at any given time. Both techniques are limited in scaling by their use of exponentially-sized arrays and the need to read each amplitude at least once.

In the field of quantum circuit simulation, an overwhelming majority of prior work focuses on computing all amplitudes through strong simulation. Here, the exponential complexity is tackled either by supercomputers (by distributing both the storage
and the processing to multiple processors~\cite{jones2019quest,DBLP:journals/corr/WeckerS14,qxSimulator2017,DBLP:journals/corr/SmelyanskiySA16}) or by dedicated data structures such as decision diagrams~\mbox{\cite{VMP:2009,zulehner2017advanced}} and Matrix Product States~\cite{DBLP:journals/qip/WangHH17}. 
However, of equal importance is efficient storage of resulting state/probability vectors and  efficient output sampling with such data representations. Thus, prior literature leaves a gap between strong simulation and the task of mimicking a physical quantum computer.

\section{Advanced Weak Simulation}
\label{sec:advanced-ws}

In this section, we develop a method for weak simulation that does not rely on
exponentially-large arrays and reduces memory blow-up in important cases.
Moreover, this method generates samples quickly after an initial precomputation.

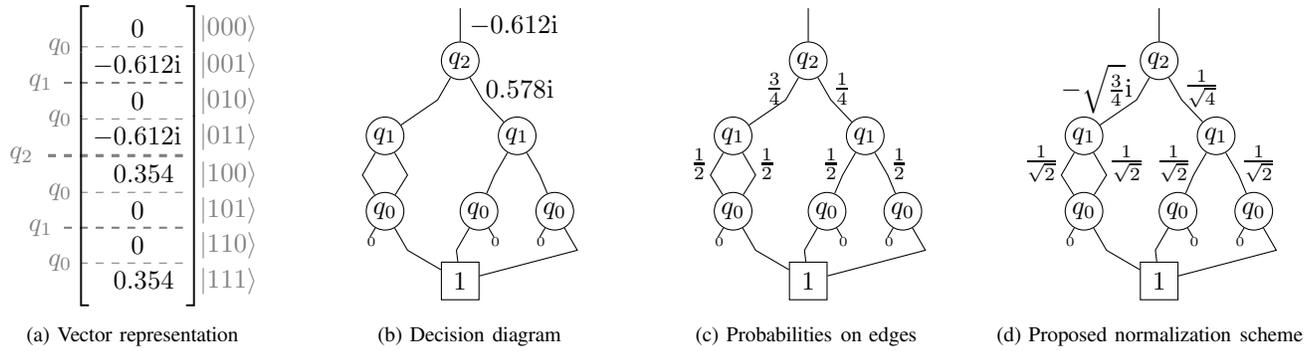
\begin{figure*}[tbp]
	\centering
	\begin{subfigure}[t]{0.24\linewidth}
		\centering
		\begin{tikzpicture}
			\matrix[matrix of math nodes, left delimiter={[},right delimiter={]}, inner xsep=0] (vector) {
				0\\
				-0.612\iu\\
				0\\
				-0.612\iu\\
				\phantom{-}0.354\phantom{\iu}\\
				0\\
				0\\
				\phantom{-}0.354\phantom{\iu}\\
			};
			
			\begin{scope}[on background layer, gray]	
				\node[right=0.7cm of vector-1-1.center] {\(\ket{000}\)};
				\node[right=0.7cm of vector-2-1.center] {\(\ket{001}\)};
				\node[right=0.7cm of vector-3-1.center] {\(\ket{010}\)};
				\node[right=0.7cm of vector-4-1.center] {\(\ket{011}\)};
				\node[right=0.7cm of vector-5-1.center] {\(\ket{100}\)};
				\node[right=0.7cm of vector-6-1.center] {\(\ket{101}\)};
				\node[right=0.7cm of vector-7-1.center] {\(\ket{110}\)};
				\node[right=0.7cm of vector-8-1.center] {\(\ket{111}\)};
				
				\draw[gray,-,dashed, very thick,shorten <= -0.6cm] ($(vector-4-1)!0.5!(vector-5-1)$) -- ++(-1.25,0) node[anchor=east] {\(q_2\)};
				
				\draw[gray,-,dashed, thick,shorten <= -0.6cm] ($(vector-2-1)!0.5!(vector-3-1)$) -- ++(-1,0) node[anchor=east] {\(q_1\)};
				\draw[gray,-,dashed, thick,shorten <= -0.6cm] ($(vector-6-1)!0.5!(vector-7-1)$) -- ++(-1,0) node[anchor=east] {\(q_1\)};
				
				\draw[gray,-,dashed, shorten <= -0.6cm] ($(vector-1-1)!0.5!(vector-2-1)$) -- ++(-.75,0) node[anchor=east] {\(q_0\)};
				\draw[gray,-,dashed, shorten <= -0.6cm] ($(vector-3-1)!0.5!(vector-4-1)$) -- ++(-.75,0) node[anchor=east] {\(q_0\)};
				\draw[gray,-,dashed, shorten <= -0.6cm] ($(vector-5-1)!0.5!(vector-6-1)$) -- ++(-.75,0) node[anchor=east] {\(q_0\)};
				\draw[gray,-,dashed, shorten <= -0.6cm] ($(vector-7-1)!0.5!(vector-8-1)$) -- ++(-.75,0) node[anchor=east] {\(q_0\)};
			\end{scope}
			\end{tikzpicture}
		\caption{Vector representation}
		\label{fig:statevectorvector}
	\end{subfigure}
	\begin{subfigure}[t]{0.24\linewidth}
		\centering
		\begin{tikzpicture}[terminal/.style={draw,rectangle,inner sep=0pt}]	
		\matrix[matrix of nodes,ampersand replacement=\&,every node/.style={draw,circle,inner sep=0pt,minimum width=0.5cm,minimum height=0.5cm},column sep={1cm,between origins},row sep={1cm,between origins}] (qmdd2) {
							\& |(m1)|$q_2$ 											\\
			|(m2a)| $q_1$	\&                                  \& |[xshift=-0.5cm](m2b)| $q_1$ 						\\
			|(m3a)| $q_0$	\& |[xshift=0.25cm](m3b)| $q_0$ 					\& |(m3c)| $q_0$ 	\\
							\& |[terminal] (t3)| $1$ 									\\
		};
		
		\draw ($(m1)+(0,0.7cm)$) -- (m1) node[right, midway]{$-0.612\iu$};
		
		\draw (m1) -- ++(240:0.6cm) node[left, midway] {} -- (m2a);
		\draw (m1) -- ++(300:0.6cm) node[right, midway] {$0.578\iu$} -- (m2b);
		
		\draw (m2a) -- ++(240:0.6cm) -- (m3a);
		\draw (m2a) -- ++(300:0.6cm) -- (m3a);
		
		\draw (m2b) -- ++(240:0.6cm) -- (m3b);
		\draw (m2b) -- ++(300:0.6cm) node[right, midway] {} -- (m3c);
		
		\draw (m3a) -- ++(240:0.4cm) node[below, xshift=0.5pt, inner sep=0,font=\tiny] {$0$};
		\draw (m3a) -- ++(300:0.6cm) -- (t3);
		
		\draw (m3b) -- ++(240:0.6cm) -- (t3);
		\draw (m3b) -- ++(300:0.4cm) node[below, xshift=0.5pt, inner sep=0,font=\tiny] {$0$};
				
		\draw (m3c) -- ++(240:0.4cm) node[below, xshift=0.5pt, inner sep=0,font=\tiny] {$0$};
		\draw (m3c) -- ++(300:0.6cm) -- (t3);
		\end{tikzpicture}
		\caption{Decision diagram}
		\label{fig:dd-statevector}
	\end{subfigure}
	\begin{subfigure}[t]{0.24\linewidth}
		\centering
		\begin{tikzpicture}[terminal/.style={draw,rectangle,inner sep=0pt}]	
			\matrix[matrix of nodes,ampersand replacement=\&,every node/.style={draw,circle,inner sep=0pt,minimum width=0.5cm,minimum height=0.5cm},column sep={1cm,between origins},row sep={1cm,between origins}] (qmdd2) {
								\& |(m1)|$q_2$ 											\\
				|(m2a)| $q_1$	\&                                  \& |[xshift=-0.5cm](m2b)| $q_1$ 						\\
				|(m3a)| $q_0$	\& |[xshift=0.25cm](m3b)| $q_0$ 					\& |(m3c)| $q_0$ 	\\
								\& |[terminal] (t3)| $1$ 									\\
			};
			
			\draw ($(m1)+(0,0.7cm)$) -- (m1) node[right, midway]{};
			
			\draw (m1) -- ++(240:0.6cm) node[midway, left]{$\frac{3}{4}$} -- (m2a);
			\draw (m1) -- ++(300:0.6cm) node[midway, right]{$\frac{1}{4}$} -- (m2b);
			
			\draw (m2a) -- ++(240:0.6cm) node[midway, left]{$\frac{1}{2}$} -- (m3a);
			\draw (m2a) -- ++(300:0.6cm) node[midway, right]{$\frac{1}{2}$} -- (m3a);
			
			\draw (m2b) -- ++(240:0.6cm) node[midway, left]{$\frac{1}{2}$} -- (m3b);
			\draw (m2b) -- ++(300:0.6cm) node[midway, right]{$\frac{1}{2}$} -- (m3c);
			
			\draw (m3a) -- ++(240:0.4cm) node[below, xshift=0.5pt, inner sep=0,font=\tiny] {$0$};
			\draw (m3a) -- ++(300:0.6cm) -- (t3);
			
			\draw (m3b) -- ++(240:0.6cm) -- (t3);
			\draw (m3b) -- ++(300:0.4cm) node[below, xshift=0.5pt, inner sep=0,font=\tiny] {$0$};
					
			\draw (m3c) -- ++(240:0.4cm) node[below, xshift=0.5pt, inner sep=0,font=\tiny] {$0$};
			\draw (m3c) -- ++(300:0.6cm) -- (t3);
		\end{tikzpicture}
		\caption{Probabilities on edges}
		\label{fig:dd-new-scheme}
	\end{subfigure}
	\begin{subfigure}[t]{0.25\linewidth}
		\centering
		\begin{tikzpicture}[terminal/.style={draw,rectangle,inner sep=0pt}]	
			\matrix[matrix of nodes,ampersand replacement=\&,every node/.style={draw,circle,inner sep=0pt,minimum width=0.5cm,minimum height=0.5cm},column sep={1cm,between origins},row sep={1cm,between origins}] (qmdd2) {
								\& |(m1)|$q_2$ 											\\
				|(m2a)| $q_1$	\&                                  \& |[xshift=-0.5cm](m2b)| $q_1$ 						\\
				|(m3a)| $q_0$	\& |[xshift=0.25cm](m3b)| $q_0$ 					\& |(m3c)| $q_0$ 	\\
								\& |[terminal] (t3)| $1$ 									\\
			};
			
			\draw ($(m1)+(0,0.7cm)$) -- (m1) node[right, midway]{};
			
			\draw (m1) -- ++(240:0.6cm) node[midway, left]{$-\sqrt{\frac{3}{4}}\iu$} -- (m2a);
			\draw (m1) -- ++(300:0.6cm) node[midway, right]{$\frac{1}{\sqrt{4}}$} -- (m2b);
			
			\draw (m2a) -- ++(240:0.6cm) node[midway, left]{$\frac{1}{\sqrt{2}}$} -- (m3a);
			\draw (m2a) -- ++(300:0.6cm) node[midway, right]{$\frac{1}{\sqrt{2}}$} -- (m3a);
			
			\draw (m2b) -- ++(240:0.6cm) node[midway, left]{$\frac{1}{\sqrt{2}}$} -- (m3b);
			\draw (m2b) -- ++(300:0.6cm) node[midway, right]{$\frac{1}{\sqrt{2}}$} -- (m3c);
			
			\draw (m3a) -- ++(240:0.4cm) node[below, xshift=0.5pt, inner sep=0,font=\tiny] {$0$};
			\draw (m3a) -- ++(300:0.6cm) -- (t3);
			
			\draw (m3b) -- ++(240:0.6cm) -- (t3);
			\draw (m3b) -- ++(300:0.4cm) node[below, xshift=0.5pt, inner sep=0,font=\tiny] {$0$};
					
			\draw (m3c) -- ++(240:0.4cm) node[below, xshift=0.5pt, inner sep=0,font=\tiny] {$0$};
			\draw (m3c) -- ++(300:0.6cm) -- (t3);
		\end{tikzpicture}
		\caption{Proposed normalization scheme}
		\label{fig:statevector-approx}
	\end{subfigure}
	\caption{Representations of the quantum state as a vector and a decision diagram}
	\vspace{-1.5em}
\end{figure*}

As outlined in Section~\ref{sec:weak-simulation}, weak simulation can be approached by first computing \emph{all} amplitudes of a state vector to determine the probability of each output basis state. Storing amplitudes and/or probabilities in exponentially-large arrays is often prohibitively expensive, but in many important cases \emph{decision diagrams} can capture quantum states and support more memory-efficient representation~\cite{DBLP:conf/esa/Samoladas08,VMP:2009,DBLP:journals/tcad/NiemannWMTD16}.
In the following, we show how such data structures can be extended to perform weak simulation. First, we provide the necessary background material on decision diagrams and, then, describe our algorithm for weak simulation. With an eye on efficient implementation, we also develop a normalization scheme for decision diagrams and combine the several ideas presented into a faithful and efficient weak simulation technique.

\subsection{Background on Decision Diagrams}
\label{subsec:decision-diagrams}

Decision diagrams have been successfully utilized with quantum circuits
for tasks such as simulation~\cite{VMP:2009,zulehner2017advanced,zulehner2019matrix},
synthesis~\cite{AP:06,niemann2018cliffordt,zulehner2017one}, and verification~\cite{WLTK:2008,burgholzer2020improved}, since they often
drastically reduce the memory needed to represent state vectors and operation matrices. 
Strong simulation approaches based on decision diagrams have recently moved into the spotlight because they can significantly outperform vector-based simulators in cases where data redundancies can be exploited---in extreme cases leading to an improvement in runtime from 30 days to 2 minutes~\cite{zulehner2017advanced}.

The main idea behind decision diagrams is to dynamically identify data redundancies and provide compaction by sharing \mbox{sub-structures}. Conceptually, the state vector is split
into two equal-sized sub-vectors. Given that the number of amplitudes is a power of two,
this process is repeated until the \mbox{sub-vectors} contain a single element only, i.e.,~one split for every qubit. Where identical sub-vectors occur in this process, this redundancy is exploited by re-using (sharing) the same \mbox{sub-structure} in the resulting decision diagram. In practice, the identification of repeated sub-vectors is performed by hashing and \mbox{bottom-up} rather than top-down. Unlike the early uses of decision diagrams in
quantum circuit simulation~\cite{VMP:2009}, we leverage edge-weighted decision diagrams~\cite{zulehner2017advanced} which provide a more powerful compression mechanism and need significantly less memory in important cases. In the following, we outline key technical details of relevant edge-valued decision diagrams to represent quantum state-vectors.

Consider a quantum system with qubits \(q_{n-1} , q_{n-2} , \ldots, q_0 \), whereby \( q_{n-1} \) represents the most significant qubit.
Then, the first \( 2^{n-1} \) entries of the corresponding state vector represent the amplitudes for the basis states with \( q_{n-1} \) set to \(\ket0\); the other entries represent the amplitudes for states with \( q_{n-1} \) set to~\(\ket1\).
This decomposition is represented in a decision diagram structure by a node labeled \( q_{n-1} \) and two successors leading to nodes representing the two sub-vectors.
By convention, the left (right) edge indicates the \mbox{0-successor} (1-successor).
The \mbox{sub-vectors} are recursively decomposed until vectors of size~\(1\) (i.e.,~complex numbers) result.
During this decomposition, equivalent sub-vectors can be represented by the same nodes---reducing the complexity of the representation.
Then, instead of having a terminal node for every distinct value in the state vector, common factors of the amplitudes are stored in the edge weights. Each amplitude
in the state vector is represented by a directed path in the decision diagram, and its
complex value can be reconstructed by multiplying the edge weights along the path.
The approach provides exponential data compression when a compact graph exhibits
a large number of directed paths.

\begin{example}
	Consider the state vector in Fig.~\ref{fig:statevectorvector}---the annotations sketch how the vector is decomposed (left) and which base state corresponds to each entry in the vector (right).
	Fig.~\ref{fig:dd-statevector} shows the corresponding decision diagram.
	Here, e.g.,~the amplitude of the state \(\ket{111}\) is accessed by following the path in the decision diagram for \(q_2 = 1\), \(q_1 = 1\), \(q_0 = 1\) and multiplying the edge weights along the path, i.e.,~\(-0.612\iu \cdot 0.578\iu \cdot 1 \cdot 1 = 0.354 \).
\end{example}

\subsection{Decision Diagrams for Weak Simulation}
\label{subsec:dd-for-weak-simulation}

Decision diagrams can compactly represent all amplitudes of a quantum state, given sufficient data redundancy.
The key idea is that instead of explicitly storing an exponential number of probabilities, we encode them in a decision diagram with the same structure as the decision diagram for amplitudes. 
This precomputation step is performed by full traversals of the decision diagram, where we compute new weights---the probabilities that the left and right successor of a node should be followed when drawing an output sample. The idea behind the subsequent sampling algorithm is to perform a randomized single-path traversal of the decision diagram from the top node, deciding randomly at each node whether to descend
to the left or right child node, based on precomputed probabilities. While physical quantum states get destroyed by quantum measurement, simulated measurement
is a read-only operation that can be repeated.

We now specify two types of probabilities computed at each node and the decision diagram traversals that calculate them:
\begin{enumerate}
	\item The \emph{downstream probabilities} are sums of probabilities of all half-paths from the current node to any terminal node.
	They are calculated by a \emph{depth-first traversal}.
	\item The \emph{upstream probabilities} are sums of probabilities of all half-paths from the root to the current node.
	They are calculated by a \emph{breadth-first traversal}.
\end{enumerate}
The runtime of the traversals is linear in the number of nodes in the decision diagram.
Once both probabilities for a node are computed, the probability to follow the left (right) successor is the product of the upstream and downstream probabilities weighted by the squared magnitude of the left (right) edge weight. 
Finally, the probability of an individual basis state is the product of all probabilities along the path.

\begin{example}
	Fig.~\ref{fig:dd-new-scheme} continues our running example and shows
the same decision diagram with the edge weights giving the probability
of choosing either successor node.
	From the outgoing edge weight of \(q_2\), it can be seen that the left successor will be chosen 3 out of 4 times, whereas the right successor will only be chosen 1 out of 4 times. That is, the resulting basis state will have $q_2 = 0$ in \SI{75}{\percent} of the samples and $q_2=1$ in \SI{25}{\percent} of the samples.
\end{example}

\subsection{Efficient Sampling}

The method described above enables sampling with reduced memory complexity in important cases, exploiting on the compactness of the representation. 
Each sample is produced by traversing a root-to-terminal path in the decision diagram. Given that for
$n$-qubit states, such paths have at most $n+2$ nodes, each output sample
is generated in $\mathcal{O}(n)$ time.
Still, the runtime of the sampling process can be improved
using a new normalization scheme for outgoing edges.

The outgoing edges of a node are often normalized by dividing both weights by the weight of the left-most edge (when $\neq 0$), and multiplying this factor
to the incoming edges as illustrated in Fig.~\ref{fig:dd-statevector}. However, we
found that in the context of sampling, it is more effective to divide by the norm of the vector containing both edge weights and adjust the incoming edges accordingly.
This normalizes the sum of the squared magnitudes of the outgoing edge weights to 1
and is consistent with the quantum semantics, where basis states \(\ket{0}\) and \(\ket{1}\) are observed after measurement with probabilities that are squared magnitudes of the respective weights.

\begin{example}
	Applying the proposed normalization scheme to Fig.~\ref{fig:dd-statevector} results in the decision diagram shown in Fig.~\ref{fig:statevector-approx}.
\end{example}

\section{Empirical Validation}
\label{sec:evaluation}

Our method for weak simulation avoids explicit exponential-sized vectors in important cases and generates output samples quickly to mimic physical quantum computers.
To evaluate it, we use a dedicated implementation of \mbox{edge-valued} decision diagrams with adaptations to quantum circuit simulation, such as the use of high-precision complex numbers, based on~\cite{zulehner2019complexvalues} (common software packages for decision diagrams are generally not usable in this context). Our empirical validation focuses on output sampling for quantum circuits that implement well-known quantum algorithms, algorithmic blocks, and applications:
\begin{itemize}
	\item The \emph{Quantum Fourier Transformation} (QFT) (denoted \enquote{qft\_\(A\)} for \(A\) qubits), 
	\item Grover's search~\cite{Gro:96} with a random oracle (denoted \enquote{grover\_\(A\)} for \(A\) qubits),
	\item Shor's algorithm to factorize integers~\cite{Sho:94} (denoted \enquote{shor\_\(A\)\_\(B\)} for factorizing~\(A\) with the coprime value~\(B\)),
	\item quantum circuits simulating the uniform electron gas~\cite{PhysRevX.8.011044} (denoted \enquote{jellium\_\(A\)x\(A\)} for size of the system \(A\times A\) on a grid), and
	\item quantum circuits provided by researchers from Google~\cite{boixo2016characterizing2} as candidates to establish quantum-computational supremacy using \mbox{controlled-Z} gates (denoted \enquote{supremacy\_\(A\)x\(B\)\_\(C\)}, representing a circuit on an \(A\times B\) surface with depth \(C\)).
\end{itemize}
Following the overall flow outlined in Section~\ref{sec:weak-simulation},
we performed strong simulation of these circuits to find the final quantum state
in the form of a decision diagram. 
Subsequently, we completed weak simulation by (1)~calculating the prefix sum and, then, conducting the sampling through binary search as described in Section~\ref{sec:weak-simulation} (\emph{vector-based}) and by (2)~using decision diagrams (\emph{DD-based}) and the normalization scheme described in Section~\ref{sec:advanced-ws}. We generated one million samples, which is common for benchmarking quantum computers today. All runs were performed on an Intel i7-7700K CPU (\SI{4.2}{\giga\hertz}) with \SI{32}{\gibi\byte} main memory (and an additional \SI{32}{\gibi\byte} of swap space) under GNU/Linux.

\begin{table}[tbp]
	\centering
	\caption{Runtime and memory for error-free sampling of 1M bitstrings}
	\label{tbl:sampling}
	\vspace{-0.25em}
	\begin{tabular}{l@{\hskip -6pt}r@{\hskip 10pt}rr@{\hskip 10pt}r@{\hskip 0pt}l>{\bfseries}r}
		\toprule
		\multicolumn{2}{c}{benchmarks}  &  \multicolumn{2}{c}{vector-based\hspace*{12pt}}  &  \multicolumn{3}{c}{DD-based}      \\
		\cmidrule(r{12pt}){1-2}
		\cmidrule(l{-3pt}r{12pt}){3-4}
		\cmidrule{5-7}
		name & qubits & size & \(t\)\,[s] & size & & {\normalfont\(t\)\,[s]}     \\ \midrule
		qft\_16 & 16 &  $2^{16}$ & 0.12 & 16 & ${} \approx 2^{4.0}$ & 0.22 \\
		qft\_32 & 32 &  $2^{32}$ &  MO & 32 & ${} \approx 2^{5.0}$ & 0.43 \\
		qft\_48 & 48 &  $2^{48}$ &  MO & 48 & ${} \approx 2^{5.5}$ & 0.63 \\[3pt]
		grover\_20 & 21 &  $2^{21}$ & 0.70 & 40 & ${} \approx 2^{5.3}$ & 0.23 \\
		grover\_25 & 26 &  $2^{26}$ & 17.91 & 50 & ${} \approx 2^{5.6}$ & 0.27 \\
		grover\_30 & 31 &  $2^{31}$ & 993.99 & 60 & ${} \approx 2^{5.9}$ & 0.29 \\
		grover\_35 & 36 &  $2^{36}$ &  MO & 70 & ${} \approx 2^{6.1}$ & 0.43 \\[3pt]
		shor\_33\_2 & 18 &  $2^{18}$ & 0.15 & 48\,793 & ${} \approx 2^{15.5}$ & 0.20 \\
		shor\_55\_2 & 18 &  $2^{18}$ & 0.16 & 93\,478 & ${} \approx 2^{16.5}$ & 0.21 \\
		shor\_69\_4 & 21 &  $2^{21}$ & 0.62 & 196\,382 & ${} \approx 2^{17.5}$ & 0.26 \\
		shor\_221\_4 & 24 &  $2^{24}$ & 3.72 & 1\,048\,574 & ${} \approx 2^{20.0}$ & 0.27 \\
		shor\_247\_4 & 24 &  $2^{24}$ & 3.81 & 1\,376\,221 & ${} \approx 2^{20.3}$ & 0.31 \\[3pt]
		jellium\_2x2 & 8 &  $2^{8}$ & 0.04 & 117 & ${} \approx 2^{6.8}$ & 0.09 \\
		jellium\_3x3 & 18 &  $2^{18}$ & 0.17 & 59\,475 & ${} \approx 2^{15.8}$ & 0.22 \\[3pt]
		supremacy\_4x4\_10 & 16 &  $2^{16}$ & 0.11 & 65\,070 & ${} \approx 2^{15.9}$ & 0.39 \\
		supremacy\_5x4\_10 & 20 &  $2^{20}$ & 0.66 & 486\,503 & ${} \approx 2^{18.8}$ & 0.82 \\
		supremacy\_5x5\_10 & 25 &  $2^{25}$ & 12.04 & 16\,779\,617 & ${} \approx 2^{24.0}$ & 4.28 \\
		\bottomrule
	\end{tabular}
	\vspace{-1em}
\end{table}

The results are presented in Table~\ref{tbl:sampling}.
The first two columns contain the name of the benchmark and the number of qubits.
The following two columns contain the size of the state vector (number of entries) that we sample from as well as the time it took to compute the prefix sum and to draw one million samples (in CPU seconds).
For DD-based sampling, similar columns show the size (number of nodes) of the state-vector decision diagram and the time it took to draw one million samples from this decision diagram.
Both approaches support faithful (error-free) weak simulation.
Moreover, the results also confirm what one might expect from complexity
analysis: If the (exponentially large) state vector can be stored in memory\footnote{For vector-based sampling, sizes above $2^{30}$ incurred swapping, the \mbox{DD-based} sampling did not require swapping in the evaluation.}, then samples can be generated using prefix sums, as described in Section~\ref{sec:weak-simulation}. Otherwise (e.g.,~for the quantum algorithms {qft\_32}, {qft\_48}, and {grover\_35}), a \emph{memory out}~(MO) occurs and this method reaches its limits.
In contrast, decision diagrams facilitate much more compact representations. Together with the sampling method proposed in Section~\ref{sec:advanced-ws}, this supports faithful and efficient weak simulation in these cases as well. Overall, our results demonstrate, for the first time, that the outputs of physical quantum computers running several \mbox{well-known} quantum algorithms can be mimicked faithfully and efficiently by simulators running on conventional computers without relying on
exponentially-sized arrays.

\section{Conclusions}
\label{sec:conclusion}

We proposed efficient methods for \emph{weak simulation}, whose output cannot be statistically distinguished from the output of (error-free) physical quantum computers. Prior literature in the field focuses on \emph{strong simulation}, i.e.,
computing the amplitudes of the final quantum state.
However, generating output samples is a distinctively different task. Hence, we outline sampling for amplitudes in an exponentially-large array and develop an alternative method
that works directly with state vectors compactly represented as decision diagrams. Both techniques (1) rely on precomputations performed in time linear in the size of input data,
and (2) generate $n$-qubit output samples in $\mathcal{O}(n)$ time. Sampling from decision
diagrams is a little slower in practice, but scales much better in terms of memory for quantum states produced by important quantum circuits. Empirical validation confirms the feasibility of our techniques and demonstrates, for the first time, a faithful mimicking of quantum computers by a simulator running on a conventional computer at scales where
exponentially-sized arrays are infeasible.

\section*{Acknowledgments}
This work has partially been supported by the LIT Secure and Correct Systems Lab funded by the State of Upper Austria.

{\footnotesize
\bibliographystyle{IEEEtran}
\bibliography{../../bib/lit_header,./new-references,../../bib/lit_misc,../../bib/lit_mymisc,../../bib/lit_myrev,../../bib/lit_others,../../bib/lit_othersrev,../../bib/lit_rev,../../bib/lit_new,../../bib/lit_adiabatic,../../bib/lit_memristor,../../bib/lit_simulation,../../bib/lit_quantum}
\eject 	
}
	
\end{document}